# IDENTIFICATION IN MODELS OF BIOCHEMICAL REACTIONS

S. A. Belbas[*][1] and Sunghee Kim[2]

___________________________________________________________________

(*)  Corresponding author.

(1)  Address:
Mathematics Department
University of Alabama
Tuscaloosa, AL. 35487-0350. USA.

Fax: (USA+) 205-348-7067
e-mail: sbelbas@gp.as.ua.edu

___________________________________________________________________

(2)  Address:
Department of Biological Sciences
University of Alabama
Tuscaloosa, AL. 35487-0344. USA.

Fax: (USA+) 205-553-6973
e-mail: skim@bsc.as.ua.edu

___________________________________________________________________

Summary

We introduce, analyze, and implement a new method for parameter identification for system of ordinary differential equations that are used to model sets of biochemical reactions. Our method relies on the integral formulation of the ODE system and a method of linear least squares applied to the integral equations. Certain variants of this method are also introduced in this paper.

Keywords: system identification, biochemical reactions, differential equations, error analysis, numerical experiments.

## 1. Introduction

There has been extensive interest in the problem of identification of parameters in mathematical models of networks of biochemical reactions. This is evident from publications like [1, 3-7].

The method we introduce and analyze in this paper is designed to exploit the fact that, in systems of ordinary differential equations with polynomial nonlinearities, the right-hand sides depend linearly on the rate constants which are to be estimated. Of course, the solution depends, in general, nonlinearly on the parameters. If the problem is formulated as fitting experimentally observed values to a trajectory obtained by solving the ODE system, then we end up with a problem of nonlinear least squares. However, if closely spaced observation instants are feasible, so that the observations can be interpreted as a perturbation (because of observation errors) of an approximation to a continuous-time trajectory, then we can formulate the estimation problem in a different way: we can ask how close the observed trajectory comes to satisfying the system of differential equations. This yields a linear least squares problem. Furthermore, the system of differential equations may be interpreted either in its original differential form, or as an integral equation. It turns out that, for the purposes of carrying out the error analysis of the related algorithms, there are certain advantages to using the integral formulation.

Thus, <u>we introduce here a method of linear least squares for nonlinear systems of ordinary differential equations with polynomial nonlinearities, based on the integral formulation of the ODE system.</u> The exact formulation of this method is given in section 2 of this paper.

An additional advantage of our method is that, since it is based on the integral formulation of the ODE system, it can also be applied, with the necessary modifications, to hereditary systems, such as those modelled by difference-differential equations and general Volterra integral equations, provided always that the equations of state dynamics (but not the solutions of those equations) depend linearly on the parameters that are to be estimated. We note that an estimation problem (of a different type) for hereditaty systems arising in biosciences was studied in Ref. [11].

The approach described in introductory textbooks relies on approximations of the exact conservation identities in a system of biochemical reactions and waiting until steady state has been achieved (or assuming that equilibrium is quickly achieved in certain of the reactions). The prototypical method of this type is the standard Lineweaver-Burk plot for Michaelis-Menten reactions, based on the Briggs-Haldane approximation.

There have been several different approaches to the identification problem in the research literature.

For linear systems of differential equations, the solution can be represented as a linear combination of exponential functions, or linear combinations of functions of the form $t^\alpha \exp(\beta t)$, and then nonlinear estimation methods, like Prony's method and other related methods, can be used. For example, ref. [8] follows the approach of nonlinear least squares estimation of the reaction rates by fitting experimental data to a linear combination of exponentials, where the coefficients in the exponents are part of the parameters to be estimated. Ref. [10] covers the case of fitting data in the case of linear ODE systems with repeated eigenvalues (which give rise to terms of the type $t^\alpha \exp(\beta t)$ in the solution). Ref. [9] follows almost the reverse approach that relies on finding a differential equation (with unknown parameters) that is satisfied by a linear combination of exponentials. Clearly, all these approaches to the identification problem are limited by the requirement of having linear differential equations. Even the simplest models of enzymatic reactions are systems of nonlinear differential equations.

Ref. [7] uses nonlinear least squares for general nonlinear systems of ordinary differential equations by using the method of multiple shooting; the examples in [7] deal with systems with right-hand side depending linearly on the unknown parameters, i.e. the same type of systems we consider in the present paper.

Ref. [1] employs a method of linear least squares for nonlinear ODE systems with linear dependence on the unknown parameters. The approach used in paper [1] amounts, in substance, to the following: a system of first-order reactions is considered; the coefficients of each term in each reaction are arranged in an array; if the system of reactions is

$$\frac{dx_i}{dt} = f_i(x), \quad x = (x_j : 1 \leq j \leq n), \quad i = 1,2,...,n$$

then each $f_i(x)$ is taken as a quadratic form in the n variables $x_1, x_2, ..., x_n$; each quadratic form has therefore $\frac{1}{2}n(n+1)$ independent coefficients, and there n such forms (one form for each equation of the system), consequently, in the formulation used in [1], there are $\frac{1}{2}n^2(n+1)$ unknown parameters; given observations $y_i(t), i = 1,2,...,n$, $0 \leq t \leq T$, the unknown coefficients are estimated in [1] by minimizing

$$\sum_i \int_0^T \left(\frac{dy_i(t)}{dt} - f_i(y(t))\right)^2 dt.$$

Since the unknown coefficients enter linearly into the differential equations, this is a linear least squares problem that can be solved by standard methods, provided the derivatives $\frac{dy_i(t)}{dt}$ can be approximated with high accuracy.

In addition to the approach of [1], it is clear that there is a need for additional work, for the following reasons:

(I). The number of unknown coefficients used in [1] is unnecessarily high; in actual equations arising from a network of n chemical reactions, the number of independent coefficients is generally less than $\frac{1}{2}n^2(n+1)$, and for large n it is considerably less than $\frac{1}{2}n^2(n+1)$.

(II). The calculation of the time-derivatives of the observation trajectories is, in general, a numerically unstable problem; if the errors in the observations themselves stay within known bounds, this does not imply any bounds for the time-derivatives of the errors, and consequently it is generally impossible to carry out an error analysis for the least squares estimation that was presented in [1].

In section 2 of the present paper, we introduce a method for the solution of the identification problem for a category of systems of biochemical reactions. Our approach differs from [1] in two respects: first, we use a formulation of the identification problem that contains a minimal set of unknown parameters, equal to the set of linearly independent reaction rates; second, we use a method that avoids taking time-derivatives of the experimental data.

For example, consider the standard simple enzymatic reaction:

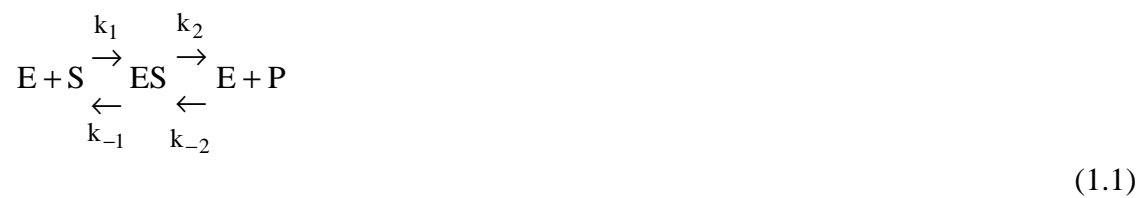

$$E + S \underset{k_{-1}}{\overset{k_1}{\rightleftarrows}} ES \underset{k_{-2}}{\overset{k_2}{\rightleftarrows}} E + P$$

(1.1)

We set

$$x_1(t) = [S](t), \; x_2(t) = [ES](t), \; x_3(t) = [E](t), \; x_4(t) = [P](t)$$

(1.2)

It is plain that the evolution of this system (without any approximations or simplifying assumptions) can be described by 2 differential equations and two algebraic equations:

$$\frac{dx_1(t)}{dt} = k_{-1}x_2 - k_1 x_1(M_1 - x_2);$$

$$\frac{dx_2(t)}{dt} = k_1 x_1(M_1 - x_2) - (k_{-1} + k_2)x_2 + k_{-2}(M_1 - x_2)(M_2 - x_1 - x_2);$$

$$x_3(t) = M_1 - x_2(t);$$

$$x_4(t) = M_2 - x_1(t) - x_2(t)$$

(1.3)

where

$$M_1 := x_2(0) + x_3(0), \; M_2 := x_1(0) + x_2(0) + x_4(0)$$

(1.4)

The solution of (1.3) really requires the solution of the two differential equations; on the other hand, for the identification problem, it may be useful to include observations of $x_3$ and $x_4$ as part of the data of the identification problem.

In the formulation we use in this paper, the identification problem for (1.3, 1.4) has 3 parameters to be estimated, namely $k_1, k_{-1}$, and $k_2$ ( we assume that the initial conditions are exactly known); in the formulation of [1], there would be 18 parameters to be estimated: an artificial variable $x_0(t) \equiv 1$ would be introduced as solution of an additional differential equation $\frac{dx_0(t)}{dt} = 0$, $x_0(0) = 1$, and the first-order terms on the right-hand sides in (1.3) would be replaced by quadratic terms $x_0 x_1, x_0 x_2$ to make the right-hand sides of all equations quadratic forms, and then the total number of differential equations would be n=3 and the total number of parameters would be $\frac{1}{2}n^2(n+1) = 18$.

The order of the reactions is not a significant factor in the choice of the method of estimation of coefficients by some version of the general idea of least squares, and thus there is no point in focusing on first-order reactions.

Consequently, in this paper, we treat systems with general nonlinearities, not just quadratic nonlinearities.

## 2. Derivation of least squares solutions.

We write the system of biochemical reactions in the form

$$\frac{dx_i(t)}{dt} = f_{i,0}(t, x(t)) + \sum_{j=1}^{N} c_j f_{ij}(t, x(t)) \text{ for } t > t_0; \ x_i(t_0) = x_{i,0}; \ i = 1, 2, \ldots, n;$$

$$x(t) := (x_1(t), \ldots, x_n(t))$$

(2.1)

The explicit dependence of the functions $f_{i,0}(t, x(t))$ and $f_{ij}(t, x(t))$ on t is relevant for cases in which some of the reactants can be considered as controlled inputs that can be modelled as exactly known functions of time. It is useful to formulate models and methods in the most general relevant form, and for this reason we have included an explicit time variable in the right-hand sides of the kinetic equations.
In order to avoid derivatives in the formulation and solution of the related least-squares problem, we write the system dynamics (2.1) in integral form:

$$x_i(t) = x_{i,0} + \int_{t_0}^{t} f_{i,0}(s, x(s)) ds + \sum_{j=1}^{N} c_j \int_{t_0}^{t} f_{ij}(s, x(s)) ds$$

(2.2)

Given observations $y_i(t)$, $t_0 \leq t \leq T$ and a real symmetric positive definite n by n matrix-valued function Q(t) (Q is postulated to be positive definite uniformly with respect to t), we seek to minimize the error E given by

$$E := \sum_{i=1}^{n} \sum_{k=1}^{n} \int_{t_0}^{T} \left[ Q_{ik}(t) \left( y_i(t_0) - y_i(t) + \int_{t_0}^{t} f_{i,0}(s, x(s)) ds + \sum_{j=1}^{N} c_j \int_{t_0}^{t} f_{ij}(s, y(s)) ds \right) \cdot \right.$$

$$\left. \cdot \left( y_k(t_0) - y_k(t) + \int_{t_0}^{t} f_{k,0}(s, x(s)) ds + \sum_{j=1}^{N} c_j \int_{t_0}^{t} f_{kj}(s, y(s)) ds \right) \right] dt$$

(2.3)

For simplicity, we set

$$g_{ij}(t) := \int_{t_0}^{t} f_{ij}(s, y(s))ds;$$

$$g_{i,0}(t) := y_i(t_0) - y_i(t) + \int_{t_0}^{t} f_{i,0}(s, y(s))ds$$

(2.4)

Setting $\dfrac{\partial E}{\partial c_m} = 0$ $(m = 1, 2, ..., N)$ yields

$$\sum_{j=1}^{N} G_{mj} c_j^* = A_m$$

(2.5)

where

$$G_{mj} = \sum_{i=1}^{n} \sum_{k=1}^{n} \int_{t_0}^{T} Q_{ik}(t) g_{ij}(t) g_{km}(t) dt,$$

$$A_m = -\sum_{i=1}^{n} \sum_{k=1}^{n} \int_{t_0}^{T} Q_{ik}(t) g_{i,0}(t) g_{km}(t) dt$$

(2.6)

and $c_j^*$ are the optimal values of the unknown coefficients.

Assuming that the expressions $g_{ij}(t)$ are linearly independent functions, the theory of Gramians implies that the matrix $G = [G_{mj}]_{\substack{1 \leq j \leq N \\ 1 \leq m \leq N}}$ is positive definite, and the solution to (2.5) can be obtained by any of the standard methods of numerical linear algebra for solving a system with positive definite coefficient matrix. If the condition of linear independence is not satisfied, then a solution based on the theory of pseudo-inverses is sought. The software package matlab has built-in functions for finding pseudo-inverses. When G is nonsingular, the solution of (2.5) can be written as

$$c^* = G^{-1} A; \quad A = [A_1 \ A_2 \ ... \ A_N]^T$$

(2.7)

A further simplification is often possible. In many cases, each term $f_{ij}(t, y(t))$ is of the form $f_{ij}(t, y(t)) = \mu_{ij} f_j(t, y(t))$ where each $\mu_{ij}$ takes one of the values 0, 1, or -1, depending on whether the reactants that are represented by the term $f_j(t, y(t))$ are absent from reactions with arrows incident to reactant $x_i$, are on the incoming (to $x_i$) side of an arrow, or are on the outgoing (from $x_i$) side of a reaction arrow, respectively. In that case, it is only necessary to calculate the terms

$$g_j(t) := \int_{t_0}^{t} f_j(s, x(s)) ds$$

(2.8)

rather than all the terms $g_{ij}$, since then

$$g_{ij}(t) = \mu_{ij} g_j(t)$$

(2.9)

In certain systems that involve fast reactions, some of the differential equations might have the form

$$\varepsilon_i \frac{dx_i(t)}{dt} = f_{i,0}(t, x(t)) + \sum_j c_j f_{ij}(t, x(t))$$

(2.10)

where $\varepsilon_i$ is a small parameter; in that case, according to the methods of differential equations with a small parameter, those differential equations can be replaced by the algebraic equations obtained by setting $\varepsilon_i = 0$, namely by

$$f_{i,0}(t, x(t)) + \sum_j c_j f_{ij}(t, x(t)) = 0$$

(2.11)

Also, algebraic equations can be obtained through establishing conservation laws for some of the variables, so that some variables, that could in principle be obtained fro differential equations, will actually be obtained from conservation equations (like the equations for $x_3(t)$ and $x_4(t)$ in (1.3). Even though, for the direct problem of finding a solution if all parameters are known, we do not need variables like $x_3(t)$ and $x_4(t)$, the inverse problem of estimating the parameters must utilize every available information, and if experimental data for variables like $x_3(t)$ and $x_4(t)$ are available, those data and the corresponding algebraic equations must be incorporated into the parameter estimation procedure.

For those equations, we simply use $f_{i,0}$ and $f_{ij}$ in place of $g_{i,0}$ and $g_{ij}$ in the formulae (2.6).

In some cases, certain individual time-instants may be of particular importance, and in that case the error functional may be modified to

$$E_1 := \sum_{i=1}^{n} \sum_{k=1}^{n} \int_{t_0}^{T} \left[ Q_{ik}(t) \left( y_i(t_0) - y_i(t) + \sum_{j=1}^{N} c_j \int_{t_0}^{t} f_{ij}(s, y(s)) ds \right) \right.$$

$$\left. \cdot \left( y_k(t_0) - y_k(t) + \sum_{j=1}^{N} c_j \int_{t_0}^{t} f_{kj}(s, y(s)) ds \right) \right] dt +$$

$$+ \sum_{i=1}^{n} \sum_{k=1}^{n} \sum_{\alpha} R_{ik}(t_\alpha) \left( y_i(t_0) - y_i(t_\alpha) + \sum_{j=1}^{N} c_j \int_{t_0}^{t_\alpha} f_{ij}(s, y(s)) ds \right) \cdot$$

$$\cdot \left( y_k(t_0) - y_k(t_\alpha) + \sum_{j=1}^{N} c_j \int_{t_0}^{t_\alpha} f_{kj}(s, y(s)) ds \right)$$

(2.12)

where the matrix $R(t)$ is real symmetric and uniformly positive definite.

In that case, the formulae (2.6) are simply modified to incorporate the discrete time-instants $\{t_\alpha\}$:

$$G_{1,mj} = \sum_{i=1}^{n} \sum_{k=1}^{n} \left[ \int_{t_0}^{T} Q_{ik}(t) g_{ij}(t) g_{km}(t) dt + \sum_{\alpha} R_{ik}(t_\alpha) g_{ij}(t_\alpha) g_{km}(t_\alpha) \right],$$

$$A_{1,m} = -\sum_{i=1}^{n} \sum_{k=1}^{n} \left[ \int_{t_0}^{T} Q_{ik}(t) g_{i,0}(t) g_{km}(t) dt + \sum_{\alpha} R_{ik}(t_\alpha) g_{i,0}(t_\alpha) g_{km}(t_\alpha) \right]$$

(2.13)

## 3. Estimation of errors.

When information about the size of the error (as a function of time) of approximating the exact state $x(t)$ by the observations $y(t)$ is available, the error in the estimation of the unknown coefficients can be estimated. If the size of the error, but not the size of the time-derivative of the error, is within given bounds, then the formulation of the problem without derivatives of $x(t)$, as in section 2 above, can show the size of the errors in the estimation of the unknown coefficients. The technique for estimating the errors is based on first-order analysis, i.e. only first-order terms in the various errors are retained. This general principle is standard in numerical analysis. In order to facilitate the exposition, we shall use the symbol " $\stackrel{(1)}{=}$ " to denote equality up to terms of first order in the errors. All the errors will be denoted by Greek letters.

We assume that the initial conditions are known exactly. We denote by x(t) the exact solution of the system (2.1) with initial conditions $x_{i,0}$, $1 \le i \le n$, and by y(t) the observations. Let the errors of observation be $\xi_i(t)$, $1 \le i \le n$, $t_0 \le t \le T$, i.e.

$$y_i(t) = x_i(t) + \xi_i(t); \quad y_i(t_0) = x_i(t_0) = x_{i,0}$$

(3.1)

Then

$$f_{i,0}(s, y(s)) \stackrel{(1)}{=} f_{i,0}(s, x(s)) - \nabla_y f_{i,0}(s, y(s)) \bullet \xi(s);$$

$$f_{ij}(s, y(s)) \stackrel{(1)}{=} f_{ij}(s, x(s)) - \nabla_y f_{ij}(s, y(s)) \bullet \xi(s)$$

(3.2)

where, for any function $\varphi(s, y)$, the gradient $\nabla_y \varphi(s, y) = [\frac{\partial \varphi(s, y)}{\partial y_k} : 1 \le k \le n]$ and

$$\nabla_y \varphi(s, y) \bullet \xi = \sum_{k=1}^{n} \frac{\partial \varphi(s, y)}{\partial y_k} \xi_k .$$

We denote by $h_{i,0}$, $h_{ij}$, $H_{mj}$, $B_m$ the terms corresponding to $g_{i,0}$, $g_{ij}$, $G_{mj}$, $A_m$ if we use the exact solution *x*(t) instead of the observations y(t), i.e.

$$h_{ij}(t) := \int_{t_0}^{t} f_{ij}(s, x(s))ds;$$

$$h_{i,0}(t) := x_i(t_0) - x_i(t) + \int_{t_0}^{t} f_{i,0}(s, x(s))ds;$$

$$H_{mj} = \sum_{i=1}^{n} \sum_{k=1}^{n} \int_{t_0}^{T} Q_{ik}(t)h_{ij}(t)h_{km}(t)dt,$$

$$B_m = -\sum_{i=1}^{n} \sum_{k=1}^{n} \int_{t_0}^{T} Q_{ik}(t)h_{i,0}(t)h_{km}(t)dt$$

(3.3)

We use the following notation for the errors associated with the above quantities:

$$g_{ij}(t) = h_{ij}(t) + \varepsilon_{ij}(t);$$

$$g_{i,0}(t) = h_{i,0}(t) + \varepsilon_{i,0}(t);$$

$$G_{mj} = H_{mj} + \eta_{mj};$$

$$A_m = B_m + \beta_m$$

(3.4)

Then

$$\overset{(1)}{\varepsilon_{ij}}(t) = -\int_{t_0}^{t} \nabla_y f_{ij}(s, y(s)) \bullet \xi(s)ds;$$

$$\overset{(1)}{\varepsilon_{i,0}} = -\xi_i(t) - \int_{t_0}^{t} \nabla_y f_{i,0}(s, y(s)) \bullet \xi(s)ds;$$

$$\overset{(1)}{\eta_{mj}} = -\sum_{i=1}^{n} \sum_{k=1}^{n} \int_{t_0}^{T} Q_{ik}(t)[g_{ij}(t)\varepsilon_{km}(t) + g_{km}(t)\varepsilon_{ij}(t)]dt;$$

$$\overset{(1)}{\beta_m} = \sum_{i=1}^{n} \sum_{k=1}^{n} \int_{t_0}^{T} Q_{ik}(t)[g_{i,0}(t)\varepsilon_{km}(t) + g_{km}(t)\varepsilon_{i,0}(t)]dt$$

(3.5)

The exact coefficients $c_j$, $j = 1, 2, ..., N$, satisfy

$$\sum_{j=1}^{N} H_{mj}c_j = B_m$$

(3.6)

whereas the optimal estimates $c_j^*$ based on the observations y(t) satisfy the system (2.5). Let the errors of estimating the coefficients be $\gamma_j$, i.e.

$$c_j^* = c_j + \gamma_j \tag{3.7}$$

Then we have

$$\sum_{j=1}^{N} G_{mj}\gamma_j \overset{(1)}{=} \beta_m \tag{3.8}$$

Consequently the first-order approximations to the errors $\gamma_j$ are found from solving (3.8) subject to (3.1), (3.2) and (3.5). These formulas can be further manipulated to estimate the norm of the vector $\gamma$ consisting of all the $\gamma_j$'s, but we omit these calculations as they follow standard procedures of numerical analysis, specifically the procedures of the branch of numerical analysis that deals with approximate solution of linear systems .

This method of estimating the error in the coefficients uses only the errors $\xi_i(t)$ but not their derivatives; by contrast, a similar estimation for the method of [1] would require knowledge about the time-derivatives $\dot{\xi}_i(t)$. An additional benefit is that, if the errors $\xi_i(t)$ are modelled as stochastic processes, the formulas of this section can be used to approximate the stochastic characteristics of the errors $\gamma_j$.

It should be noted that the estimates above have practical significance; (3.8) allows us to estimate the errors in terms of observable and computable quantities. The terms $G_{mj}$ and $\beta_m$ in (3.8) are computable functions of the observations $y_i(t)$, $i = 1,2,...,n$, $0 \leq t \leq T$.

We prove:

Theorem 3.1. Assume that all functions $f_{ij}$ and $f_{i,0}$ are continuously differentiable in all their variables, conditions for existence and uniqueness of solutions of (2.1) are satisfied, and the matrix Q is continuous in t and positive definite for all t in $[t_0, T]$. Further, we assume that the n-dimensional-vector-valued functions $\mathbf{f}_j(t, x(t)) \equiv \left[f_{ij}(t, x(t))\right]_{1 \leq i \leq n}$ are linearly independent. Then the system (3.6) has a unique solution, which must necessarily equal the exact values of the reaction rates.

<u>Proof:</u> The linear independence of $\{\mathbf{f}_j(t,x(t)): 1 \le j \le N\}$ is equivalent to the linear independence of $\{\mathbf{h}_j(t,x(t)): 1 \le j \le N\}$ where $\mathbf{h}_j(t,x(t)) \equiv [h_{ij}(t,x(t))]_{1 \le i \le n}$ ; indeed, the linear independence of the $\mathbf{f}_j$'s is tantamount to the condition that

$$\sum_j \mu_j f_{ij}(t,x(t)) = 0 \ \forall i \ \forall t \text{ implies } \mu_j = 0 \ \forall j,$$ with an analogous condition for the linear independence of the $\mathbf{h}_j$'s ; now, the conditions $\sum_j \mu_j f_{ij}(t,x(t)) = 0 \ \forall i \ \forall t$ and

$\sum_j \mu_j h_{ij}(t,x(t)) = 0 \ \forall i \ \forall t$ are equivalent, the first is obtained by differentiating the second, and the second by integrating the first. Under the stated assumptions, the matrix **H** with elements $H_{jm}$ defined in (3.3) is positive definite, since it is a generalized Gramian matrix, specifically $H_{jm} = \int_{t_0}^{T} \mathbf{f}_j^T \mathbf{Q} \mathbf{f}_m \, dt$ , and positive definiteness follows by the same type of argument as for ordinary Gramians; for completeness, we outline this argument. It is clear that $\sum_{j,m} H_{jm} \mu_j \mu_m \ge 0$ for every N-dimensional vector $\boldsymbol{\mu}$, i.e. **H** is positive semi-definite; it remains to show that, for nonzero $\boldsymbol{\mu}$, $\sum_{j,m} H_{jm} \mu_j \mu_m > 0$ . Take an arbitrary nonzero N-dimensional vector $\boldsymbol{\mu}$ ; then

$$\sum_{j,m} H_{jm} \mu_j \mu_m = \sum_{i,k} \int_{t_0}^{T} Q_{ik}(t) \left( \sum_j \mu_j h_{ij}(t) \right) \left( \sum_m \mu_m h_{km}(t) \right) dt$$

(3.9)

Since Q is positive definite, the vanishing of $\sum_{j,m} H_{jm} \mu_j \mu_m$ would amount to the vanishing of $\sum_j \mu_j h_{ij}(t)$ for all i and all t, which is impossible by the linear independence condition. Thus **H** is positive definite, and consequently it is nonsingular. The unique solution of (3.6) must give the true values of the coefficients $c_j$ since the true coefficients certainly minimize the quadratic error. ∎

Under the continuity assumptions of theorem 3.1, the rank condition on **f** is preserved under sufficiently small perturbations of x(t), since the rank condition can be algebraically expressed as the non-vanishing of certain determinants made up of elements of **f**, and inequalities are preserved under small perturbations of continuous functions. Under those condition, the solution of (2.5) depends continuously on the coefficients of the matrix $[G_{jm}]$, and thus continuously on the errors $\xi_i(t)$. Therefore, we have:

<u>Theorem 3.2.</u> Under the conditions of theorem 3.1, for sufficiently small observation errors $\xi_i(t)$, the system (2.5) has a unique solution $\{c_j^*: 1 \leq j \leq N\}$, and $c_j^* \to c_j \; \forall j$ as $\max_{t_0 \leq t \leq T, 1 \leq i \leq n} |\xi_i(t)| \to 0$.
∎

Theorem 3.2 establishes the validity of the linear least squares formulation of the problem, and justifies using this formulation instead of other nonlinear formulations that have been proposed in the literature.

The effect of the errors $\{\gamma_j: 1 \leq j \leq N\}$ in the estimation of the coefficients $c_j$ on the solution x(t) can be estimated by the techniques used in proving the continuous and differentiable dependence of the solution of an ODE system on the right-hand side of the system.

We have formulated the results above for the case of the system of differential equations (2.1), but analogous results also hold for a system consisting of a mixture of differential equations and algebraic equations.

The effect of the estimation errors on the solution of system (2.1) can be assessed in the following way:

Let x(t) be the solution of (2.1), and $x^*(t)$ the solution of the system obtained from (2.1) by substituting $c_j^* = c_j + \gamma_j$ in lieu of $c_j$ and retaining the same initial conditions. Define

$$\delta x_i(t) = x_i^*(t) - x_i(t)$$

(3.10)

Then δx satisfies

$$\frac{d}{dt}\delta x_i(t) \stackrel{(1)}{=} \sum_{k=1}^{n} \left\{ \frac{\partial f_{i,0}(t,x(t))}{\partial x_k} + \sum_{j=1}^{N} c_j \frac{\partial f_{ij}(t,x(t))}{\partial x_k} \right\} \delta x_k(t) +$$

$$+ \sum_{j=1}^{N} \gamma_j f_{ij}(t,x(t));$$

$$\delta x_i(t_0) = 0$$

(3.11)

Therefore, we consider the system of differential equations

$$\frac{d}{dt} z_i(t) = \sum_{k=1}^{n} \left\{ \frac{\partial f_{i,0}(t,x(t))}{\partial x_k} + \sum_{j=1}^{N} c_j \frac{\partial f_{ij}(t,x(t))}{\partial x_k} \right\} z_k(t) +$$

$$+ \sum_{j=1}^{N} \gamma_j f_{ij}(t,x(t));$$

$$z_i(t_0) = 0$$

(3.12)

whose solution gives the first-order approximation to the errors $\delta x_i(t)$. The solution of (3.12) can be represented in terms of the $n \times n$ fundamental matrix $\Phi(t,s)$ that solves

$$\frac{\partial}{\partial t} \Phi_{ij}(t,s) = \sum_{k=1}^{n} \left\{ \frac{\partial f_{i,0}(t,x(t))}{\partial x_k} + \sum_{m=1}^{N} c_m \frac{\partial f_{im}(t,x(t))}{\partial x_k} \right\} \Phi_{kj}(t,s);$$

$$\Phi_{ij}(s,s) = \delta_{ij} \equiv \begin{cases} 1 \text{ for } i = j \\ 0 \text{ for } i \neq j \end{cases}$$

(3.13)

Then, by the variation of constants formula,

$$z_i(t) = \sum_{j=1}^{N} \gamma_j \sum_{k=1}^{n} \int_{t_0}^{t} \Phi_{ik}(t,s) f_{kj}(s,x(s)) ds$$

(3.14)

i.e., the first-order approximations to the errors in the calculation of the solution x(t) are linear combinations of the errors in the estimation of the reaction rates with coefficients given in (3.14).

## 4. Models with rational functions.

Many biochemical models involve not just polynomial nonlinearities, but also nonlinearities involving rational functions. Such rational functions arise from quasi-steady-state assumptions in some of the reactions. Examples of biochemical models with rational function nonlinearities are well known; see, for example, [3] and other works in the same general area.

For the purposes of solving parameter identification problem, we take the general model of such reactions in the form

$$\frac{dx_i(t)}{dt} = \frac{P_i(t, x(t))}{R_i(t, x(t))}; \quad x_i(t_0) = x_{i,0} \tag{4.1}$$

where

$$P_i(t, x(t)) = p_{i,0}(t, x(t)) + \sum_j c_j p_{ij}(t, x(t)), \quad R_i(t, x(t)) = r_{i,0}(t, x(t)) + \sum_j c_j r_{ij}(t, x(t)) \tag{4.2}$$

and we have used, for notational convenience, the same set of coefficients $c_j$ in both the numerators $P_i(t, x(t))$ and the denominators $R_i(t, x(t))$; clearly, this does not entail any loss of generality, since, if different coefficients appear in the numerators and the denominators, we can use the union of the sets of all coefficients to produce the set $\{c_j\}$, and if one of the $c_j$'s appears only in a numerator then we take the corresponding $r_{ij}$ equal to zero, and similarly if one of the $c_j$'s appears only in a denominator.

Many of the models presented in [3] and other references have the general form

$$\frac{dx_i(t)}{dt} = \sum_j \prod_k \frac{a_{ijk}(x_k(t))^{q_{ijk}}}{(K_{ijk})^{q_{ijk}} + (x_k(t))^{q_{ijk}}} f_{ijk}(t, x(t))$$

where each $f_{ijk}$ is polynomial in $x$ (in many cases, each $f_{ijk}$ is a constant, which can be incorporated into $a_{ijk}$); these systems can easily be transformed into the form (4.1) after doing the straightforward algebra.

Given observations $y_i(t)$, $t_0 \leq t \leq T$, we formulate the identification problem as the minimization of the error

$$E := \int_{t_0}^{T} Q_{ik}(t)[R_i(t,y(t))\dot{y}_i(t) - P_i(t,y(t))][R_k(t,y(t))\dot{y}_k(t) - P_k(t,y(t))]dt$$

(4.3)

In this case, it is generally impossible to avoid using the time-derivatives $\dot{y}(t)$ and still maintain a problem that is linear in the unknown coefficients $\{c_j\}$. Other than this observation, the problem is analogous to the problem of section 2 above, where now, instead of (2.4), we have

$$g_{i,0}(t) = r_{i,0}(t,y(t))\dot{y}_i(t) - p_{i,0}(t,y(t)),$$

$$g_{ij}(t) = r_{ij}(t,y(t))\dot{y}_i(t) - p_{ij}(t,y(t))$$

(4.4)

The optimal solution $c^*$ for the vector of unknown coefficients is found from equations (2.5) through (2.7), but utilizing (4.4) instead of (2.4).

**5. Numerical examples.**

We present and discuss two examples of parameter estimation in systems of differential equations of the type that arises in biochemical reactions.

Example 5.1. We solve a problem by two methods, to show the advantages and limitations of different methods.

For this example, we use the same differential equations as in the single numerical example of [1], but with different initial conditions. The system is

$$\frac{dx_1(t)}{dt} = x_2(t)$$
$$\frac{dx_2(t)}{dt} = -x_1(t) + x_2(t) - x_2(t)x_3(t)$$
$$\frac{dx_3(t)}{dt} = k((x_1(t))^2 - x_3(t)) + (x_1(t))^2$$

(5.1)

The correspondence with the notation of section 2 is given by

$$f_{1,0}(t,x) = x_2; \; f_{2,0}(t,x) = -x_1 + x_2 - x_2 x_3; \; f_{3,0}(t,x) = x_1^2;$$
$$f_{11}(t,x) = f_{21}(t,x) = 0; \; f_{31}(t,x) = x_1^2 - x_3; \; c_1 = k$$

We obtained a numerical approximation to the solution of this system over $0 \le t \le 20$ with k=1 and initial conditions $x_1(0) = 2, x_2(0) = 4, x_3(0) = 7$. The solution was carried out by using the adaptive Runge-Kutta-Fehlberg method of orders (4, 5) of the matlab package, with tolerance $10^{-8}$. The solution is plotted in figure 5.1. We also solved the same initial value problem with artificial "observations" $y(t) = (y_1(t), y_2(t), y_3(t))$ obtained from x(t) by adding Gaussian noise with mean zero and standard deviation 0.06; the pseudo-random Gaussian numbers were generated by the matlab function randn. The initial values of y were set equal to the initial values of x , all other values of y were perturbed by noise; this was done for consistency with the theoretical situation described in sections 2 and 3, but the algorithm works just as well if the initial values of y are also perturbed. The "observations" y(t) are plotted in figure 5.2.

**Insert Figures 5.1 and 5.2 here.**

We used the method we have introduced in section 2 of this paper to estimate the single parameter $c_1 \equiv k$, with Q equal to the identity matrix. The appropriate integrals that appear in the formulae of section 2 were approximated by trapezoidal sums. The resulting estimate of k was 1.006814190785238. We also solved the same estimation problem by minimizing the functional

$$E_d := \int_0^{20} \left[ \frac{dy_3(t)}{dt} - c_1((y_1(t))^2 - y_3(t)) - (y_1(t))^2 \right]^2 dt$$

(5.2)

(this second method is, in substance, the method used in [1]), and we obtained an estimated value of 0.9985184428089214 for the parameter k. (Of course, these numerical values are not exactly reproducible, i.e. repeating the same numerical experiment would not generally produce precisely the values 1.006814190785238 and 0.9985184428089214 for the estimates of k, because every implementation of the randn function produces a different set of pseudo-random numbers; the general properties of the algorithms, over a wide range of implementations, are reproducible.) For the numerical implementation of the least squares method for minimizing the functional above, we approximated the derivatives by averaged backward and forward difference-quotients (except at the beginning and the end of the time-period, where we used only forward and only backward approximations, respectively). The integrals were again approximated by trapezoidal sums.

Two questions arise in connection with the numerical results reported above:

(1). The performance of the method of section 2 can be theoretically anticipated and justified on the basis of the estimates of section 3. Is there a mathematical justification for the success of the second method, i.e. the method of minimizing (5.2)?
(2). It appears that the first method overestimates the wanted parameter, whereas the second method underestimates the same parameter. Is this a coincidence, or is it systematic? If it is systematic, is there a mathematical explanation of this behaviour of the two methods?

The answer to question (1) is that, for observation errors that can be modelled as independent random variables in general, and Gaussian random variables in particular, it is possible to estimate the size of the errors in the numerical approximation of time derivatives (such as the derivative $\frac{dy_3(t)}{dt}$), with high probability. For the case at hand, suppose that $y_3(t_k) = x_3(t_k) + \xi_3(t_k)$ where $t_k$ are the subdivision points of the Runge-Kutta-Fehlberg method, and the $\xi_3(t_k)$ are independent random variables with mean zero and standard deviation $s$. The derivative $\frac{dy_3(t_k)}{dt}$ is numerically approximated by

$$Dy_3(t_k) \equiv \frac{1}{2}\left[\frac{y_3(t_{k+1})-y_3(t_k)}{h_{k+1}} + \frac{y_3(t_k)-y_3(t_{k-1})}{h_k}\right]$$

(5.3)

where $h_k \equiv t_k - t_{k-1}$, $h_{k+1} \equiv t_{k+1} - t_k$, and the corresponding error of observation is

$$\vartheta_3(t_k) \equiv \frac{1}{2}\left[\frac{\xi_3(t_{k+1})-\xi_3(t_k)}{h_{k+1}} + \frac{\xi_3(t_k)-\xi_3(t_{k-1})}{h_k}\right].$$

It is easy to see that the $\vartheta_3(t_k)$ are random variables with mean zero and standard deviation $s_\vartheta = \frac{s}{\sqrt{2}}\left[h_k^{-2} + h_{k+1}^{-2} - (h_k h_{k+1})^{-1}\right]^{1/2}$. Then, by Chebyshev's inequality, the absolute values of $\vartheta_3(t_k)$ are not greater than $as_\vartheta$ with probability not less than $1-a^{-2}$, for each positive number a. The $\vartheta_3(t_k)$ are generally not independent from each other, so additional work is needed for carrying out estimates analogous to those of section 2; we do not carry out such estimates in this paper. (We note that the individual difference-quotients $\frac{\xi_3(t_{k+1})-\xi_3(t_k)}{h_{k+1}}$ would be mutually independent if the observation errors $\xi_3(t_k)$ constituted a discrete version of a Wiener process.) Nevertheless, the errors of the estimation process that uses numerical approximations to the time-derivatives of the observations (and thus also of the process of [1]) can be estimated, since there is an estimate (with high probability) of the errors in the derivatives of the observations, and the performance of this second algorithm can be theoretically established. In the case of random variables with specific distributions, better estimates can be given; for example, for a Gaussian random variable with mean 0 and standard deviation s, the absolute values of the random variable do not exceed $3.3s$ with probability 0.999 (cf. any statistical tables or software). As an illustration, if $h_k = h_{k+1} = \frac{1}{15}$ and the errors are Gaussian with mean 0 and s=0.06, then, with probability 99.9%, the observation errors in the numerical approximation of the time-derivatives do not exceed

$\frac{(3.3)(0.06)(15)}{\sqrt{2}} \approx 2.1001$; by contrast, the observation errors in the trajectories themselves do not exceed $3.3(0.06) = 0.198$ with probability 99.9%. In other words, there is an amplification factor, equal to the expression

$\frac{1}{\sqrt{2}}\left[h_k^{-2} + h_{k+1}^{-2} - (h_k h_{k+1})^{-1}\right]^{1/2}$ , when we go from the observation errors of a

trajectory to the observation errors of the time-derivative of the trajectory. For numerical confirmation, we have plotted the exact time derivative $\frac{dx_3(t)}{dt}$ in figure 5.3, and the numerical approximation to the time-derivative of the observations, i.e. to $\frac{dy_3(t)}{dt}$, in figure 5.4. The variable magnitude of the noise in $\frac{dy_3(t)}{dt}$ for different time-instants is due to the adaptive nature of the Runge-Kutta-Fehlberg method that generates unequal time-steps over the interval $[t_0, T]$.

### Insert Figures 5.3 and 5.4 here.

It must be pointed out that the success of the second method in this example is due to the specific way of generating noisy observations in computer simulations, and not to the intrinsic mathematical properties of this method.

The answer to the second question is that, for the system (5.1) with the specified initial conditions, the method of section 2 generally overestimates the parameter k, and the second method of minimizing the functional (5.2) generally underestimates k. We have plotted, in figure 5.5, the estimates based on the two methods for different values of the noise intensity s; in order to emphatically exhibit the tendencies of the two methods, we have carried out numerical experiments for higher values of s (even though those values of s would probably not arise in practice, since they are high compared to the exact solution x(t)) and the results are plotted in figure 5.6. All numerical experiments corresponding to figures 5.5 and 5.6 were carried out with higher accuracy Runge-Kutta-Fehlberg methods, with tolerance $10^{-10}$.

### Insert Figures 5.5 and 5.6 here.

The mathematical explanation of this behaviour of the two methods as follows:

Both methods lead to equations of the form $Gk^* = A$ for the best estimate $k^*$ of the parameter k. In the method of section 2 of this paper, the calculation of G involves squares of integrals of $(x_1^2 - x_3)$ and the calculation of A involves products of integrals of $x_1^2$ and integrals of $(x_1^2 - x_3)$; in the solution with the given initial conditions, it follows from the plots of $x_1(t)$ and $x_3(t)$ that the integral terms appearing in the evaluation of A exceed the terms appearing in the evaluation of G, resulting in greater positive error terms in the evaluation of A than for G, i.e. G is underestimated relative to A, and consequently the solution $k^* = \dfrac{A}{G}$ is an overestimation of the true value of k. Almost the reverse situation occurs in the estimation of k based on minimizing (5.2): now the evaluation of G involves integrals of the squares of the time-derivatives of the observations, whereas the evaluation of A involves only the first powers of the same time derivatives; because of the amplification factor in the observation errors of the time-derivatives, as explained above, now G is overestimated relative to A, and correspondingly $k^* = \dfrac{A}{G}$ is now an underestimation of the true value of k. These are, of course, statements of a probabilistic nature, i.e. the remarks above hold with high probability; the performance of the two algorithms for isolated cases may be different, but, statistically, the two algorithms will generally behave as described above. Further, more precise and detailed mathematical analysis of these algorithms is possible, but it is beyond the scope of this paper.

These remarks show the importance of having more than one methods of estimation: it can happen that, for intrinsic theoretical reasons, one method will overestimate and another method will underestimate the same parameters, and the best information can be obtained by combining different methods.

Example 5.2. We solve the parameter estimation problem for system (1.3) by using the method of section 2. We write the system (1.3) in the form

$$\frac{dx_1(t)}{dt} = k_{-1}x_2(t) - k_1 x_1(t) x_3(t)$$

$$\frac{dx_2(t)}{dt} = k_1 x_1(t) x_3(t) - (k_{-1} + k_2) x_2(t) + k_{-2} x_3(t) x_4(t)$$

$$x_2(t) + x_3(t) - M_1 = 0$$

$$x_1(t) + x_2(t) + x_4(t) = M_2$$

(5.4)

Purely for illustration purposes, we take the values

$$k_1 = 0.5, k_{-1} = 0.1, k_2 = 0.2, k_{-2} = 0.05, M_1 = 10, M_2 = 30,$$
$$x_1(0) = M_2, x_2(0) = 0, t_0 = 0, T = 0.6$$

(5.5)

The solution of (5.4, 5.5) by the Runge-Kutta-Fehlberg method of orders 4 and 5 and error tolerance $10^{-10}$ is shown in figure 5.7; simulated observations, with Gaussian noise representing observation errors, with standard deviation s=0.05, are shown in figure 5.8.

**Insert figures 5.7 and 5.8 here.**

For levels of noise s=0.01 and s=0.05, we ran several simulations, using a new implementation of a random number generator for each simulation. In each case, we estimated the parameters $k_1, k_{-1}, k_2, k_{-2}$ by the method introduced in section 2, for the case of a mixed system consisting of differential equations and algebraic equations. The results are presented in tables 5.1 and 5.2. In order to display the effects of the noise intensity on the estimated quantities, we have also run simulations for an artificially high value of noise intensity, s=0.5, and the results of the estimation algorithm of section2 are shown in table 5.3.

___
_

Table 5.1.
Estimated values of $k_1$, $k_{-1}$, $k_2$, $k_{-2}$ with standard deviation of noise s=0.01, for several implementations of perturbing the exact solution with random Gaussian noise.

```
  4.998008302368606e-001
  9.761668391621734e-002
  2.274603920887190e-001
  9.495490013044612e-001

  4.998633602896566e-001
  1.021501608325987e-001
  1.993046710459362e-001
  4.862406145992351e-002

  4.997860360902254e-001
  9.993499081230826e-002
  2.279753458817672e-001
  1.019418291336919e+000

  4.998338043962294e-001
  1.000051663282115e-001
  2.006866641657069e-001
  7.432037059495521e-002
```

___
_

___

Table 5.2.
Estimated values of $k_1$, $k_{-1}$, $k_2$, $k_{-2}$ with standard deviation of noise s=0.05, for several implementations of perturbing the exact solution with random Gaussian noise.

```
 5.008324977746548e-001
 8.937186969471209e-002
 2.578292894698573e-001
 1.761892573433419e+000

 5.019384924021661e-001
 1.067722840957128e-001
 1.995640462116934e-001
 4.654596768994646e-002

 4.987623148699284e-001
 1.136988621945352e-001
 1.908575076643431e-001
-1.474666305511221e-001

 5.010874095747399e-001
 1.006612458190666e-001
 1.968021219920316e-001
-6.880599418326444e-002
```

___

___
_

Table 5.3.
Estimated values of $k_1$, $k_{-1}$, $k_2$, $k_{-2}$ with standard deviation of noise s=0.5, for several implementations of perturbing the exact solution with random Gaussian noise. In these examples, the noise intensity was unrealistically high, in order to demonstrate the tendency of the algorithm with regard to each of the parameters $k_1$, $k_{-1}$, $k_2$, $k_{-2}$.

```
 4.916901418707019e-001
 3.756398219297605e-002
 9.562022462392560e-002
-3.879838670144778e+000

 4.912248594457358e-001
 8.857586807596517e-002
 2.030314461144840e-001
-2.518386099165492e+000

 5.046747666778225e-001
 8.872621635783919e-002
 3.117811070805918e-001
 2.581000288804610e+000

 5.055736821439726e-001
 1.277121537174511e-001
 1.339052492316449e-001
-1.414326187907284e+000
```

___
_

In order to assess the effect of the noise, as distinct from the effect of the numerical approximations (numerical solution of the ODE system, numerical evaluation of integrals, numerical solution of a linear system) associated with the method of section 2, we also calculated the best estimates for the reaction rates for the case s=0 (no noise). The results for s=0 were

$$k_1^* = 4.998965680260972\text{e-}001$$
$$k_{-1}^* = 9.999783401168126\text{e-}002$$
$$k_2^* = 1.998188364671055\text{e-}001$$
$$k_{-2}^* = 4.359438660520458\text{e-}002$$

It is seen from these results that, in terms of closeness of estimated values to exact values, the parameter with the highest numerical value, viz. $k_1 = 0.5$, gives the best results, while $k_{-2}$ gives the worst results; this is to be expected, since $k_{-2}$ is multiplied by the term $x_3 x_4$, which is small because $x_4$ is small (cf. figure 5.7). As the noise intensity increases, naturally the performance for $k_{-2}$ gets worse, whereas the performance for $k_1$ remains robust for high values of noise. The other two parameters, $k_{-1}$ and $k_2$, behave in ways that are intermediate between the two extreme cases, $k_1$ and $k_{-2}$. It must be noted that these remarks apply to the degree of closeness of estimated values to exact values, which, of course, is not the same with sensitivity of the solution of (5.4) with respect to errors in the estimation of the reaction rates; for example, a marked error in the estimation of $k_{-2}$ will have a small effect of the solution because of the smallness of the term $x_3 x_4$, so that the relevant quantity is not just $k_{-2}$ but rather the monomial $k_{-2} x_3 x_4$.

**Figure legends**

Figure 5.1. Solution of (5.1) with the stated initial conditions.

Figure 5.2. Simulation of observations obtaining by perturbing the solution x(t) of (5.1.) with Gausian noise with mean 0 and standard deviation s=0.06.

Figure 5.3. Derivative $\frac{dx_3}{dt}$ for the solution shown in figure 5.1.

Figure 5.4. Numerical approximation to the derivative $\frac{dy_3}{dt}$ for the simulated observations shown in figure 5.2.

Figure 5.5. Estimated values of k by two methods, for different values of the standard deviation of the noise.

Figure 5.6. Tendencies of two methods to, respectively, overestimate and underestimate k. In order to emphatically display those tendencies, the values of the standard deviation s of the noise have been extended to high values.

Figure 5.7. Numerical solution of system (5.4) with data given by (5.5).

Figure 5.8. Perturbation of the solution shown in figure 5.7 by adding Gaussian noise with mean 0 and standard deviation s=0.05.

Figure 5.1.

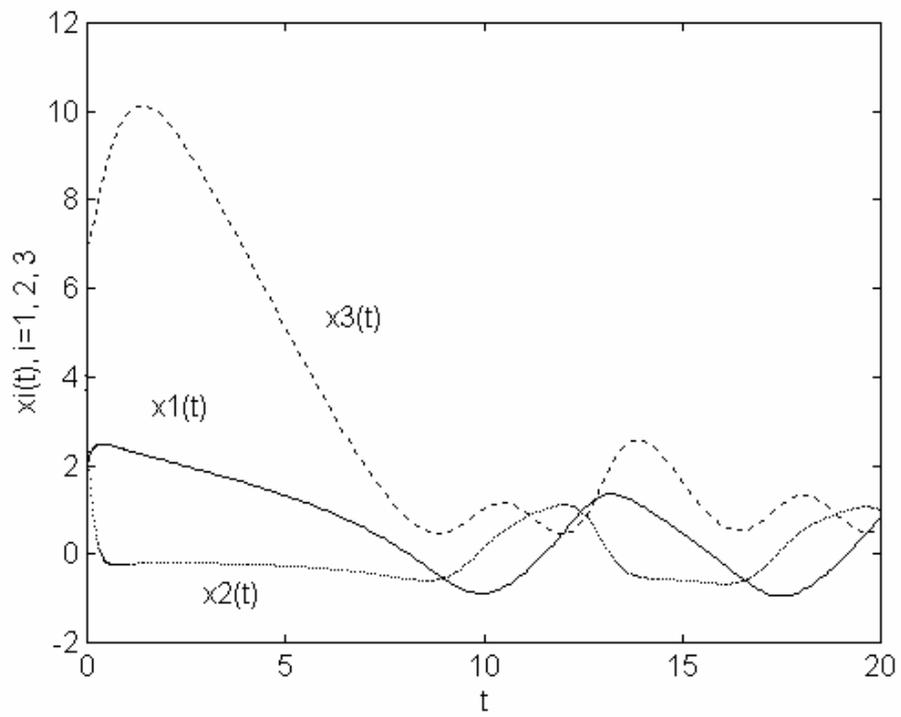

Figure 5.2.

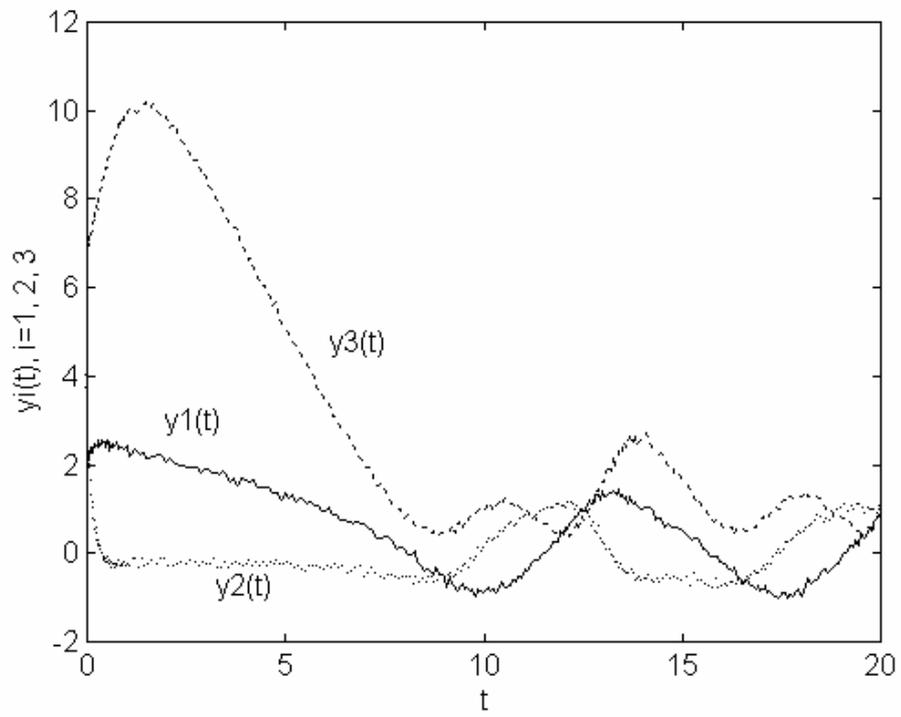

Figure 5.3.

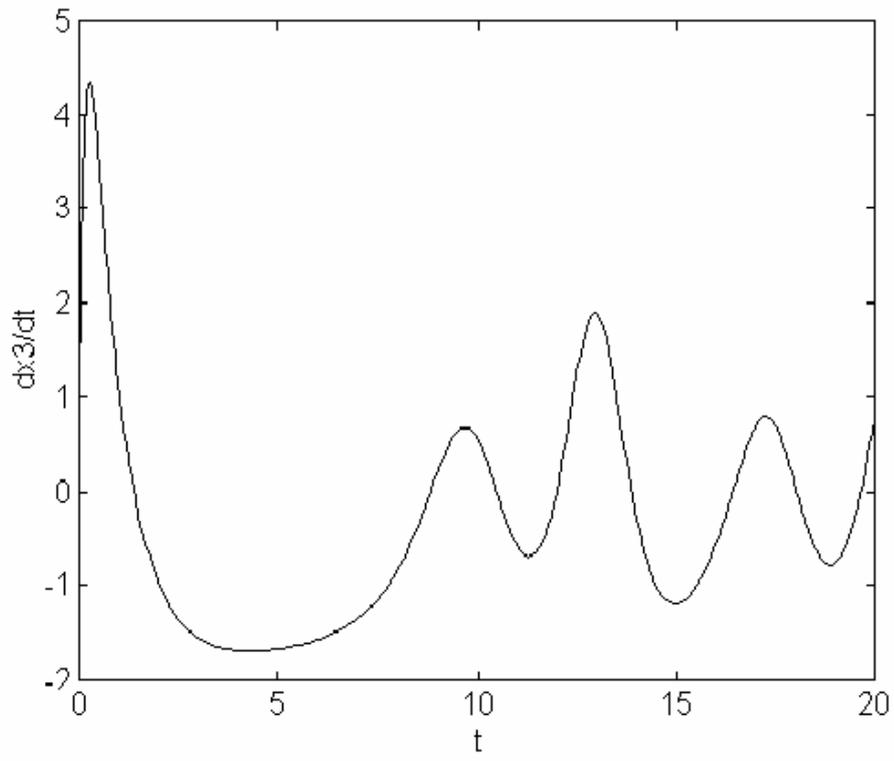

Figure 5.4.

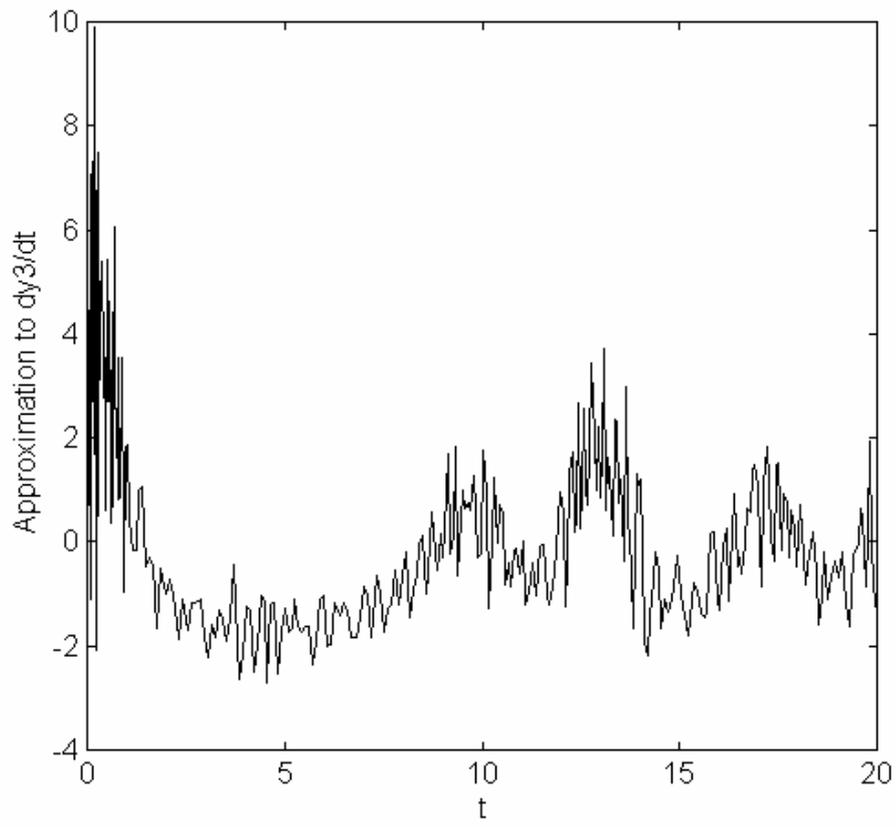

Figure 5.5.

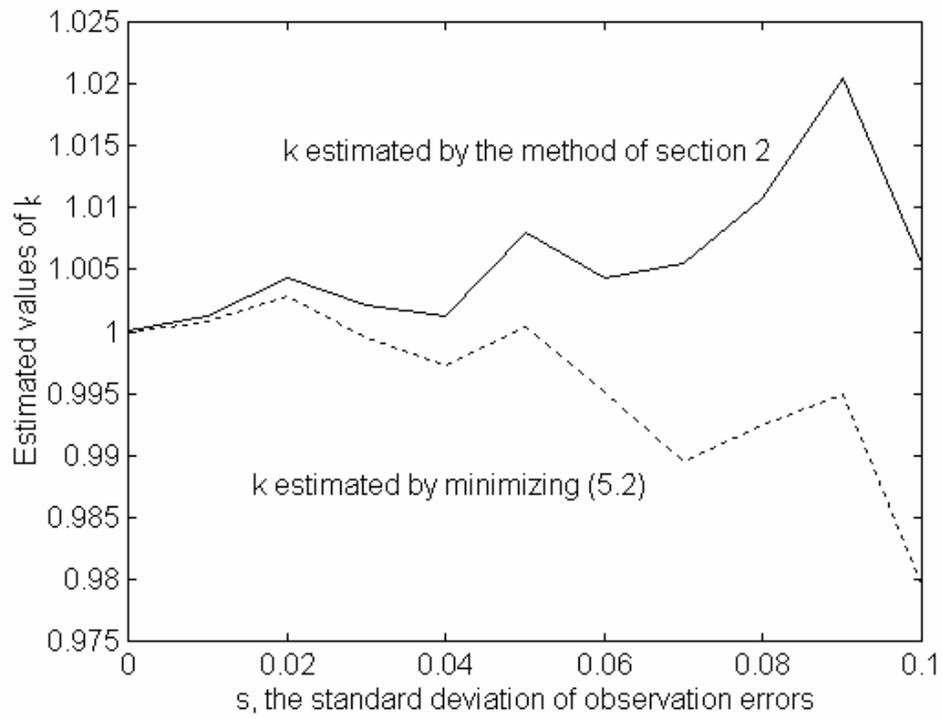

Figure 5.6.

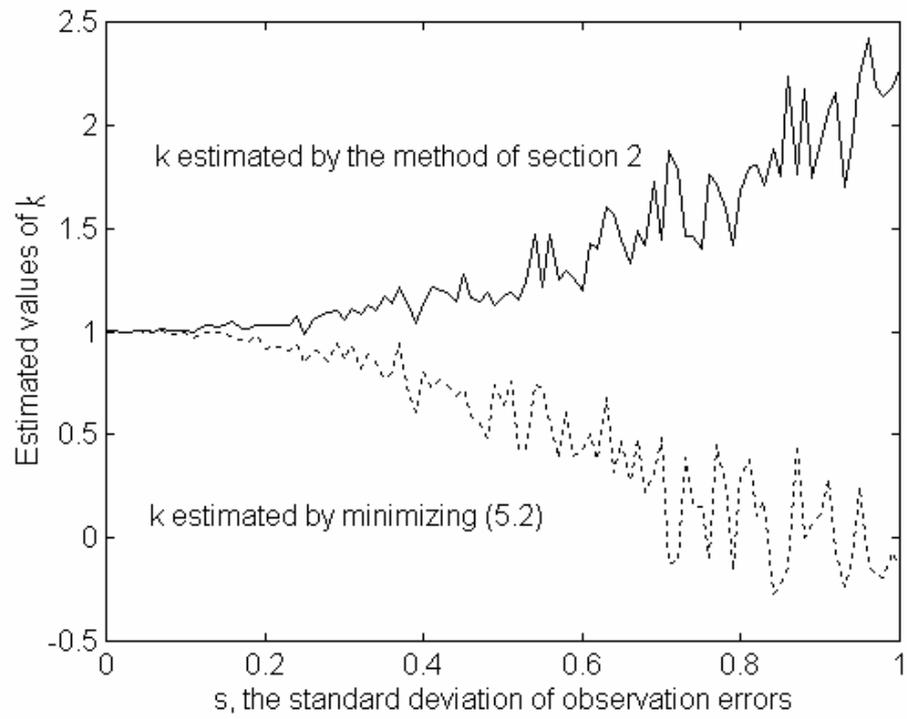

Figure 5.7.

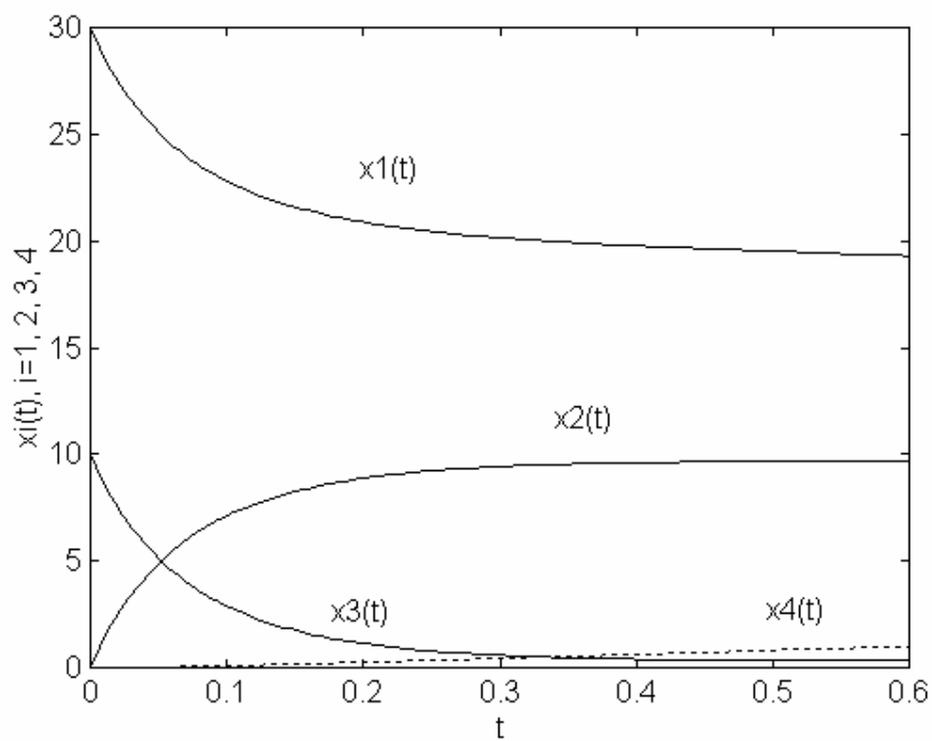

Figure 5.8.

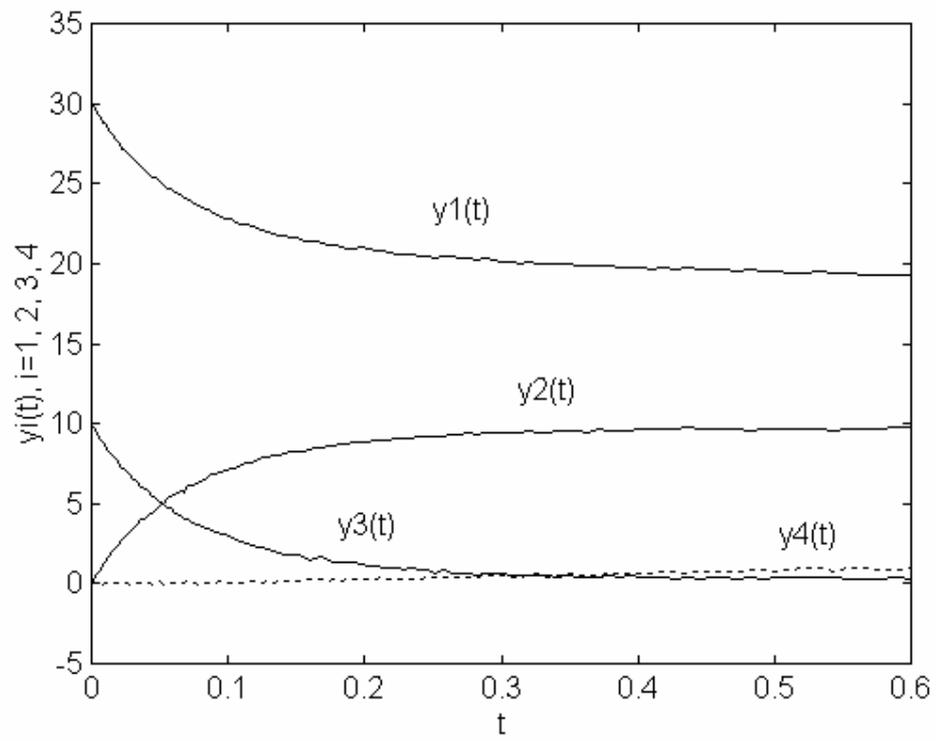